\documentclass{article}

\usepackage{graphicx}
\usepackage{siunitx}
\usepackage{float} 
\usepackage{natbib} 
\usepackage{authblk} 
\usepackage{url}
\usepackage{amsmath}
\usepackage{amsfonts}

\title{BBE: Simulating the Microstructural Dynamics of an In-Play Betting Exchange\\ via Agent-Based Modelling\footnote{Copyright \copyright\/ 2021, D.\ Cliff.}}

\author{Dave Cliff}

\date{}

\affil{\small Department of Computer Science, University of Bristol, Bristol BS8 1UB, U.K.\\{\tt csdtc@bristol.ac.uk}; {{\sc Orcid} 0000-0003-3822-9364}}

\begin{document}

\maketitle
\vspace*{-1cm}
\begin{abstract}
This paper describes the rationale for, and design of, an agent-based simulation model of a contemporary online sports-betting exchange: such exchanges, closely related to the exchange mechanisms at the heart of major financial markets, have revolutionized the gambling industry in the past 20 years, but gathering sufficiently large quantities of rich and temporally high-resolution (sub-second interval) data from real exchanges -- i.e., the sort of data that is needed in large quantities for leading-edge machine learning techniques such as Deep Learning -- is often very expensive, and sometimes simply impossible; this creates a need for a plausibly realistic synthetic data generator (SDG), which is what the simulation described here is intended to provide. The simulator, named the {\em Bristol Betting Exchange}\/ (BBE),
is intended to become a common platform, a data-source and experimental test-bed, for any/all researchers studying the application of artificial intelligence (AI) and machine learning (ML) techniques to issues arising in betting exchanges; and, as far as I have been able to determine, BBE is the first of its kind: a free open-source agent-based simulation model consisting not only of a sports-betting exchange, but also a minimal simulation model of racetrack sporting events (e.g., horse-races or car-races) about which bets may be made, and a population of simulated bettors who each form their own private evaluation of odds and place bets on the exchange before and -- crucially -- during the race itself (i.e., so-called ``in-play'' betting) and whose betting opinions change second-by-second as each race event unfolds. BBE is offered as a proof-of-concept simulator system that enables the generation of large high-resolution data-sets for automated discovery or improvement of profitable strategies for betting on sporting events via the application of AI/ML and advanced data analytics techniques. This paper offers an extensive survey of relevant literature and explains the motivation and design of BBE, and presents brief illustrative results. In a companion paper, I and my co-authors describe three independent implementations of BBE that each serve different audiences, for which the source-code is being released on GitHub, and show comparative results.      
\end{abstract}


\section{Introduction}

Like it or not, humans have been gambling for fun and profit (and sometimes for ruin too, sadly) for a very long time: \cite{hombas_baloglou_1995} document the record of gambling in ancient Egyptian and ancient Greek cultures. Right now, more than four thousand years later, according to  \cite{PRNews_2020} global revenues for the gambling industry are commonly estimated to be moving up in excess of US\$500billion per year.

For the vast majority of the past four thousand years, gambling was a fairly simple and technology-free process: if two people who like to bet (i.e.\ two {\em bettors}) denoted by B1 and B2 take opposing views on the outcome of some uncertain future event, such as the toss of a coin or the roll of a die, then they might agree a sum of money, the {\em wager} or {\em stake}, and then when the event's outcome is known either B1 will pay to B2, or B2 pays B1. Colloquially, we might say that B1 bets on one outcome (e.g. ``heads", i.e., the tossed coin landing with the heads-side showing) while B2 bets on the complementary outcome (``tails''), but in the technical language that we'll use in this paper, we'll instead say that B1 {\em backed}\/ heads (i.e., bet that heads {\em would}\/ be the outcome) while B2 {\em layed}\/ heads (i.e., bet that heads {\em would not}\/ be the outcome).
 
If our two bettors B1 and B2 decide that instead of betting against each other they would prefer to take their chances with a commercial bookmaker (i.e., a ``bookie''), either an individual or an organization, the same language applies: B1 could wager \$100 with the bookie on a particular horse winning a race, or a particular team winning a soccer match, or any other uncertain event, in which case B1 is {\em backing}\/ the specified outcome, while the bookie is {\em laying}\/ the outcome. For very many years, this was most people's commonplace experience of commercial betting: a bettor backs an outcome with a bookie, accepting the odds (also known as the {\em price}) that the bookie specifies at the time, and the bookie lays the bet. In principle, it was possible for an individual bettor to make a lay bet: some bookmakers would offer a price (usually after being given time to work out the odds) on an outcome that is the complement-set of a single specific outcome -- e.g. ``{\em I bet \$100 that any team except for Team X will win the Championship this year}'' -- in such circumstances the bettor is laying Team X and the bookie is backing Team X, but for most people in everyday betting circumstances the usual implicit understanding is that bettors back and bookies lay.

Since the dot-com boom of the late 1990's, the worldwide betting industry has transformed from being focused almost entirely on traditional bookmaking as previously practiced for hundreds of years, to one in which the dominant practice revolves around bettors buying and selling bets on betting exchanges, and particularly on the internationally successful UK-based betting exchange {\em Betfair} (see \cite{betfair_2021}), which is widely credited with being the disruptive innovator in this space, and which rapidly grew to huge financial success. Other commercial betting exchanges include \cite{betdaq_2021} and \cite{smarkets_2021}. 

The key innovation in Betfair was recognising that the existence of a population of bettors with varying and opposing views, such that the various possible outcomes of a sporting event each attract some number of bettors willing to back that outcome, and some other number of bettors willing to lay the outcome (i.e., because they want to back some opposing outcome), is essentially the same situation as in a financial market where there are some number of traders interested in buying an asset, and some number of traders wanting to sell the asset. The reason why the services offered by Betfair and similar platforms are described as betting {\em exchanges}\/ is because they bring backers and layers together to identify counterparties to a bet, and give all participants a shared summary view of the distribution of bets for a particular event, in a manner very similar to how most major financial exchanges bring together potential buyers and sellers, give them a summary view of the overall market supply and demand for some tradeable asset, and allow traders in the market to identify counterparties willing to transact at a price that both parties consider fair.

Briefly, almost all current electronic financial exchanges in the global markets operate what is known as a {\em Limit Order Book}\/ (LOB) which aggregates and anonymizes the orders received from the various traders active in the market for a particular asset. Different traders will often have different prices at which they are willing to transact, and the quantities they have available to sell, or are willing to buy, will also differ: all of this is captured in the LOB, which is usually updated in real-time after each new order (or cancellation of a prior order) is received from a trader, with the updated LOB then immediately published to all traders simultaneously. In contemporary financial markets the LOB for a liquid asset will be changing and updated many times per second as traders issue new orders or cancel existing ones, but at any one instant the LOB displays a rich array of information showing all the prices at which traders in the market are willing to buy or sell, and the total quantities demanded or supplied at each of those prices: a trader looking at a snapshot of the LOB can see the state of the entire market at that time. Sometimes there is enough information in a single LOB snapshot for a trader to confidently make short-term predictions about which way transaction prices are about to move, e.g. because the LOB shows an imbalance between supply and demand in the market; but more generally it can be informative to view a sequence of changes in the LOB over some time period, i.e.\ as a time-series of snapshots, to identify and predict longer-term trends in supply, demand, and prices.  

\begin{figure}
\begin{center}
\includegraphics[scale=0.35]{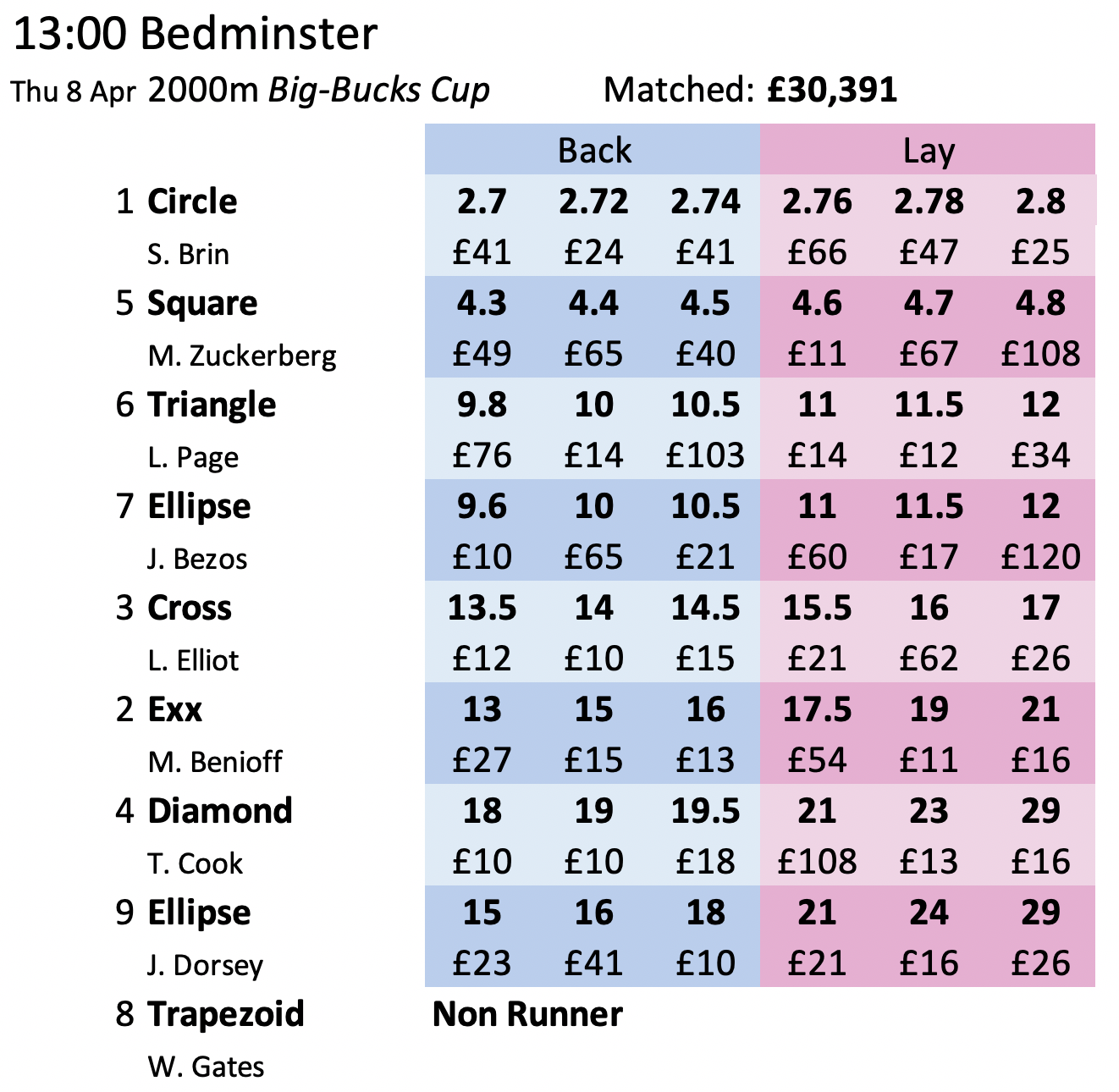}
\end{center}
\caption{A typical {\em grid}\/ (i.e., tabular) display for the ``market'' (i.e.\ the order-book) for a particular race-event on which bets are being made by bettors backing and/or laying the event. This fictional event is a nine-horse race at the Bedminster track, run over 2000m, taking place on 8th April, and starting at 1pm. The name of this race is the Big-Bucks Cup, and purely coincidentally all the jockeys have names quite similar to various famous wealthy people, while the horses are all named after simple shapes. Data for each horse is displayed on a row of the table, arranged from top-to-bottom in order of lowest-to-highest odds (so the horse at the top of the table is the current {\em favourite}, i.e.\ the one that the current betting currently indicates is most likely to win).  The far-left column shows the race-number for each horse, and the column to the immediate right of that shows the name of that horse and its jockey. The remaining cells of the table show an upper number in boldface which is the betting odds (or `prices'), as a decimal value, the total return if the bet succeeds, explained further in Section~\ref{sec:background}, and below that the total amount of money currently staked at that price -- the total stake could be from a single bet, or the sum of multiple bets. The blue-shaded half of the table shows data from back-bets, and the pink-shaded half shows data from lay-bets. The best back prices are those with the highest odds; the best lay prices are those with the lowest odds: this grid shows the best three back prices, and the best three lay prices, for each horse.  One horse, Number 8 {\em Trapezoid}, ridden by Wilma Gates, is a non-runner.} 
\label{fig:betlob}
\end{figure}

Betfair's key innovation was to create something very like a LOB, adapted to matching backers and layers in a betting market: at the heart of a betting exchange for a particular event is a data structure which is referred to as the {\em market} for that event, which is a direct analogue of the LOB in a financial market. Betting exchanges offer markets for many types of event, but in the rest of this paper for simplicity I'll limit the discussion to track-racing events. The market will typically be displayed to a bettor as a rectangular table or {\em grid}\/ of cells, with each competitor in the race allocated one row on the grid. A betting exchange's market for an event is split between backs and lays, arranged in order of goodness-of-odds, so that each cell in the grid displays a specific odds along with the total amount wagered at those odds: Figure~\ref{fig:betlob} shows an illustrative example. However, the grid display only shows a small number (typically three, sometimes one) of the best prices available to back and lay a particular competitor. 

Crucially, a betting exchange is not acting like a traditional bookie: it is not carrying risk of losing its own money by laying a customer's back, or backing a customer's lay: instead, it is acting merely as a centralized meeting and matching service for bettors to seek and identify willing counterparties with whom to bet. The exchange will typically generate revenue by charging a small percentage fee (e.g. 5\%) as commission on winning bets; losing bets are commonly not charged any commission.

An additional innovation made popular by Betfair, and the central focus of BBE, is so-called {\em in-play}\/ betting, where customers can continue to trade on the exchange, laying or backing bets on the outcome of a sporting event such as a horse race or a football match after the event has started, right up until some previously-specified end-time such as the first horse passing the winning post or the final whistle in a soccer match. In-play betting is also sometimes also referred to as {\em in-running} or {\em in-race} or {\em in-game} betting.  BBE has been specifically designed to model in-play betting, and it is (as far as I can determine) unique in that regard: an issue discussed further in Section~\ref{sec:litrev}. One key aspect of in-play betting that BBE has been designed to explore is the opinion dynamics within the population of bettors; i.e., the extent to which the opinions of some bettors in the market for an event have their opinions (and hence their subsequent bets) affected by the distribution of money on the market for that event, and by any sudden changes in that distribution -- because the distribution of money in the market for an event gives insight into the collective opinions, the overall sentiment, of the population of bettors active in that market. Relevant research literature on opinion dynamics is discussed further in Section~\ref{sec:litrev}.     

\begin{figure}
\begin{center}
\includegraphics[scale=0.4]{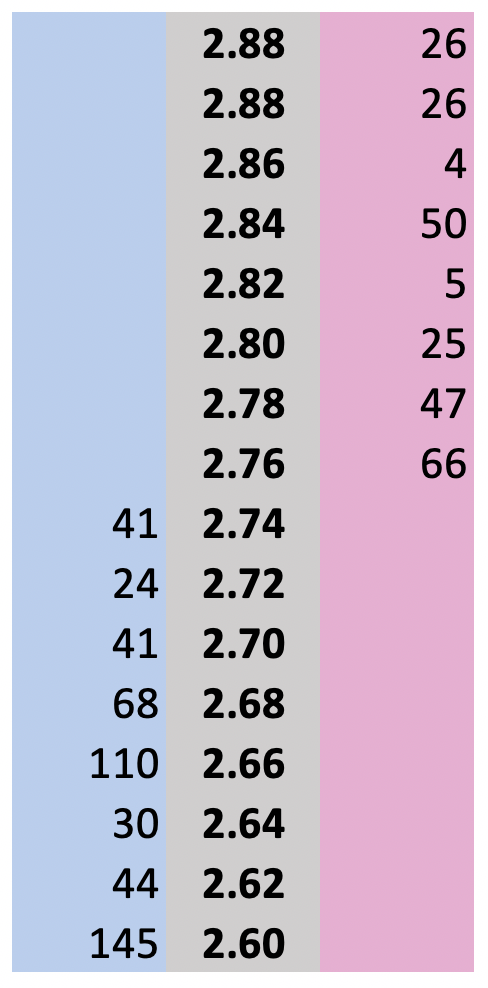}
\end{center}
\caption{A typical {\em ladder}\/ (i.e., linear) display for the ``market'' (i.e.\ the order-book) for a particular event or competitor in a betting exchange. This is for the same fictional event as was illustrated in Figure~\ref{fig:betlob}, and shows the segment of the ladder for the horse named {\em Circle}, centered on the best back and lay odds. The central  column shows the odds, and the left and right columns show the {\em liquidity}\/ (i.e., total amount of unmatched bets staked) for backs and lays, respectively. The full ladder would typically run from a low price of 1.01 to a maximum of 1000 or higher, although toward the extremes many of the cells would be expected to be empty (i.e., have no money staked at that price). The ladder view offers the advantage of showing full market {\em depth} (i.e., the complete array of odds and the liquidity at each price) to a bettor.} 
\label{fig:betladder}
\end{figure}

With a betting exchange publishing details of its in-play market for the various possible outcomes of a particular event comes the opportunity for bettors to risk their money on derivative bets, i.e.\ to not bet directly on the actual outcome of the event, but instead to wager a stake on price-movements within the market while the event is taking place: this is sometimes referred to as {\em trading}, to distinguish it from betting on the event itself, and is described in more detail in Section~\ref{sec:background}. Traders on betting exchanges often find the limited summary data in the grid-view of the market to be too restrictive, and opt instead for an interface that displays the {\em ladder}\/ for each competitor, a linear display of every available odds/price, and the {\em liquidity} (total sum of staked amounts) at each price, as illustrated in Figure~\ref{fig:betladder}.

From its foundation in June 2000, the Betfair exchange grew rapidly, to the point that by 2010 it was widely reported to be processing more transactions per day than all European stock exchanges combined \cite{roy_2010}, and various companies now offer technology tools and services for automated betting on the Betfair exchange, under the proposition that an automated betting system can be more profitable and less time-consuming than a human user manually placing bets and monitoring events. 

For discussion of the growth and impact of betting exchanges such as Betfair see \cite{davies_etal_2005,cameron_2009}; and \cite{kucharski_2017}; for a populist introduction to betting on Betfair, see \cite{houghton_2006}; and for two entertainingly populist accounts of the mathematics of gambling see \cite{poundstone_2005} and \cite{thorp_2017}. The collections edited by \cite{hausch_etal_1994,vaughanwilliams_2003,vaughanwilliams_2009,hausch_ziemba_2008}; and \cite{rodriguez_etal_2017} provide detailed academic analyses of various aspects of these developments and related aspects of the sports betting industry: the review by \cite{smith_vaughanwilliams_2008} is particularly relevant to the topics of this paper. 

Devising profitable automated betting strategies is a labor-intensive activity requiring significant expertise in the design/development phase, and potentially needing access to very large amounts of Betfair exchange data, i.e.\ time-series of various betting markets on which strategies can be tested. Betfair does sell such data, but charges premium fees which can be prohibitive for non-professional betting-strategy developers, thereby erecting a major barrier to entry. A primary motivation for the design and development of BBE was to create a source of reliable synthetic data that could be used to explore the application and refinement of artificial intelligence (AI) and machine learning (ML) methods to Betfair-style in-play betting-exchange datasets, thereby facilitating replacement of the skilled human betting-strategy designer with automated analysis, search and optimization processes: this is returned to in Section~\ref{sec:furtherwork} later in this paper. The motivation for BBE is not solely for saving money on data fees: for some data-intensive machine learning approaches, the amounts of data required simply exceed the amount that is plausibly available. For example, there might be some killer-app ML approach that will provably learn a hugely profitable bet-trading strategy, so long as you can provide it with the last 100 years of trading data from that betting exchange: because no exchange has been running for long enough, relying on real data from an exchange is not a viable approach. If, instead, reliable synthetic data can be used to train the ML system to be a profitable automated bettor, that could then be tested/evaluated on the real data from the actual exchange, and the automated system could then be deployed to generate large pots of money.

This problem, of AI/ML requiring original data at impracticable scales, is now increasingly addressed by the creation of {\em synthetic data generators} (SDGs; see, e.g., \cite{elemam_etal_2021}). For the purposes of this paper, I'll define a {\em constructive}\/ SDG as a model that can create new data-sets which preserve the original data's key statistical features, and for which the ground-truths are known and explainable -- that is, for which we know and control the causal mechanistic interactions that led to the generation of the data. This contrasts with what I'll refer to as a {\em reproductive}\/ SDG which takes some set of original data and which generates new data that reproduces the original's statistical profile, but without having anything to say about the underlying causal mechanistic interactions that generated (and explain) the underlying ground truths. The ground-truth requirement means that we are not reliant on inference to determine what set of world conditions gave rise to the data: instead we have a record of exactly what caused the data to be as they are. That is, BBE can be considered as a constructive SDG for a contemporary sports exchange, concentrating on the generation of plausible synthetic data-sets for in-play betting on track-racing events, such as horse-races, from an agent-based model (ABM) in which the causal mechanistic interactions are explicit, known, and under full control of the experimentor. 

This paper concentrates on design issues in bringing BBE up to operational capability: this involves creating agent-based models of race events -- the competitors and the bettors, and the betting-exchange matching engine, all of which are described here. Three independent implementations of BBE have been produced and are being released as open-source on {\em GitHub}, as documented in more detail by \cite{cliff_etal_2021_emss}, which is a follow-on to this paper, and which summarises the implementations developed by \cite{hawkins_2021_MEng,keen_2021_MEng}, and \cite{lausoto_2021_MEng}. We intend in future to fine-tune the implementations such that the statistical properties of the data synthesized by BBE are in the best possible alignment with those of data from one or more existing commercially-operated betting exchanges: this is discussed further in Section~\ref{sec:furtherwork}.  

BBE involves three primary components: the minimally-simulated track-racing events that the bettors bet on; the betting exchange where backs and lays can be posted and matched; and the population of bettors themselves. These are discussed in Sections~\ref{sec:racesim}, \ref{sec:exchange}, and~\ref{sec:bettors} respectively. Section~\ref{sec:results} then presents brief illustrative results from the current BBE. Section~\ref{sec:furtherwork}  talks about future work that can build on the current BBE implementations available on GitHub; and conclusions are drawn in Section~\ref{sec:conclusion}. Before that, Section~\ref{sec:background} gives further background information, and Section~\ref{sec:litrev} reviews related work in the literature.

\section{Background: Betting and Financial Markets}
\label{sec:background}

For completeness, we should first note here that an alternative way of organizing betting is as a {\em pari-mutuel}\/ or {\em totalizer} system where the bookie pools all the money bet on a particular event such as a horse-race into a single ``pot'' of cash, deducts the bookie's fee from the pot, and then when the outcome of the event is known the money remaining in the pot is divided among those bettors who backed the correct outcome, with bettor payouts made pro-rata to the amounts wagered by each winning bettor, modulo the odds computed by the bookie working from the distribution of bets wagered. This style of betting is not modelled in BBE, because betting exchanges operate via a different mechanism.

It seems a fair assumption that readers of this paper do not require an explanation of the basics of how a track-race event operates, how it is set up and run: all of the aspects of a real track-race event that are captured in our model are described in precise detail in Section~\ref{sec:racesim}. 

Currently BBE models only markets for {\em win bets} (i.e., betting that a particular competitor will win the race, or not): real betting exchange also offer additional markets for any particular event, such as {\em place bets} (e.g., betting that a particular competitor will come in the top $N$, where $N$ often depends on the size of the field, i.e.\ the number of competitors running) and {\em each way} bets (i.e., a combined pair of a win bet and a place bet, on the same race); the addition of such supplementary markets in BBE, enabling study of the interplay between win and other markets, is an obvious avenue for future work. 

The basics of betting on an exchange such as Betfair have already been introduced in the previous section, but the relationship between betting exchanges and financial exchanges does bear further discussion, to facilitate the review of related literature that follows in Section~\ref{sec:litrev}. In the text that follows, I first give a brief illustrative description of the core operations at the heart of a contemporary electronic financial market, and then go on to discuss the issue of modelling the traders in a financial market, as a preamble to the issue of how best to model the bettors in a betting market. 

The operation of a LOB-based financial exchange is best explained by an example. Consider the market for a fictitious asset with the unique identifying ``ticker" code of  XYZ, and say there are three sellers: seller S1 has issued an order to the exchange to sell 3 units of XYZ for an asking-price (the {\em ask}) of \$100 each; seller S2's order shows that she wants to sell 2 units for \$90; and seller S3 also wants to sell 2 units for \$90, so her order is the same as S2's, but happens to have been sent to the exchange after S2's was. The exchange's matching engine for XYZ would show the ask-side of the XYZ LOB as an ordered set of (price, quantity) pairs arranged in best-to-worst price order: ((\$90,4), (\$100,3)) -- note that the identity of the sellers is hidden, and that the orders from sellers S2 and S3 have been aggregated into a total quantity of 4 available at the single best-ask price of \$90. Continuing the same example, if there are four buyers in the market, B1 to B4, and their current orders (their {\em bids}) are 2 at \$75, 1 at \$80, 1 at \$75, and 3 at \$70, respectively, then the bid-side of the LOB would again aggregate and anonymize the orders and arrange them in best-to-worst price order, so the bid-side would be ((\$80,1), (\$75,3), (\$70,3)). If a Buyer B5 then came in with a bid-order of (\$100,5), the exchange's LOB matching engine would automatically match B5's bid with the asks received from S2 and S3, and with one of the units offered at \$100 by S1, and would remove those orders from the ask-side LOB (leaving it showing as ((\$100,2)) -- i.e., the residual of S1's ask) and would also notify the traders concerned that their orders had been matched with a counterparty, at which point their transactions go into a {\em clearing} process that handles the necessary transfers of ownership and of funds. If instead B5's order had been a bid of (\$100,3), the matching engine would have consumed all of S2's two items offered at \$90 and one of the two offered at the same price by S3; in this case S2's order is matched before S3's, because S2's order arrived at the exchange before S3's: that is, the exchange's engine is here operating a {\em time-priority}\/ matching process. For more detailed discussion of LOB dynamics, see \cite{gould_etal_2013_LOBs,abergel_etal_2016} and \cite{bouchaud_etal_2018}. 

The market for an event in a betting exchange such as Betfair is a close correlate of the LOB: instead of traders issuing orders to the financial exchange that each signal a desired direction (buy/sell), price, and quantity, bettors on a betting exchange issue orders that each state a direction (lay/back), the odds (a.k.a. price), and the stake (i.e., an amount of money). 

In this paper, as on most major betting exchanges, all odds will be expressed as decimals (potential total returned), rather than using fractional or American ``moneyline'' representations. For example, where a successful bet with a \$1 stake generates winnings of \$10 plus the original stake returned, for a total of \$11, the decimal representation of the odds is 11 (fractional: 10/1; moneyline: +1000); similarly, where a successful bet with a \$5 stake returns winnings of \$1 plus the original stake returned for a total of \$6, the decimal odds are 1.2 (fractional: 1/5; moneyline: -500). 

For any one sports-betting market, the betting exchange will have a matching engine running continuously that receives orders from the bettors, and aggregates and anonymizes them in a way directly analogous to the LOB matching-engine in a financial exchange: rather than show for each price the total quantities offered and bid-for by sellers and buyers respectively, the betting-exchange market shows, for each odds-price backed or laid, the total amount of currency staked at that price; and, as in a financial market, when it comes to matching orders from one counterparty to another, time-priority matching is commonly used. Further details of the implementation of the BBE matching engine are given in Section~\ref{sec:exchange}; before that, in Section~\ref{sec:litrev}, we review the literature and public-domain software for financial-exchange simulations, which have both proven to be useful resources in the construction of BBE. 

As was mentioned in the Introduction, bettors may also engage in derivative bets or ``trading", i.e.\ gambling on price movements within the market for a particular event, rather than directly on the outcome of the event itself.  For example, one well-known trading approach is ``dobbing", where the initial letters d-o-b stand for double-or-bust.  Dobbing requires that a bettor stakes money on the hope that the odds for a particular outcome will fall by 50\% or more in the course of the race: the bettor backs the outcome at the initial high odds, the starting price, and then lays the outcome once its market price has fallen by 50\% or more i.e., once its odds have shortened, reflecting the market's collective opinion that the event is now more likely than was than initially assumed. This can be illustrated with a simple horse-racing example: an initial \$10 staked on a back bet at odds of 22.0 would give the bettor \$220 if the horse wins, and would lose the \$10 stake if the horse does not win; but if before the end of the race the horse's price falls to 11.0 and the bettor places a doubled stake of \$20 as lay bet on that horse (known as {\em laying off}), then if the horse doesn't win the payout on the lay is the \$20, giving the bettor a profit of \$10 after deduction of the initial back stake; but if the horse does win then the bettor has to pay out \$220 on the lay, and wins \$220 on the back, for a net zero (or ``bust"); the two possible outcomes are either zero or positive profit (\$10 in this example), and hence over a sequence of dobbings the expected accumulated profit is positive. So, by repeated successful dobbing, a bettor can make money regardless of the outcome of the races involved. Of course there will be some additional costs such as exchange fees, and if the price of a horse that a bettor has dobbed fails to shift in the desired direction then the bettor can be left nursing a loss, but the fact remains that the dobbing bettor's strategy is not focused on whether the horse wins or loses, but instead is a gamble on the movements of the horse's price within the market as the event progresses. Readers familiar with trading in financial markets will recognise dobbing as an instance of {\em hedging}, and the opportunities for hedging on betting exchanges are explained in detail by \cite{axen_cortis_2020}.

Just as one would expect an agent-based model of a financial exchange to include some number of simulated traders issuing bid and ask orders to the exchange, thereby populating its LOB, and eventually to enter into trades with one another; so in BBE we need not only a simulation of the betting exchange's matching engine to give a continuously updating order book for that race-event's market, but also some number of simulated bettors to each form an opinion on the likely outcome of the event, and to place back and lay bets accordingly: some of the bettors should be betting directly on the outcome of the event, but others might be engaged in trading activities.

As is discussed in more detail in the next section, while there is a rich seam of literature on simulating traders for financial markets, there seems to be almost no literature on simulating bettors for betting markets. Relevant literature is reviewed in more depth in the next section, but before that it is useful to explicitly tease out the extent to which a group of bettors placing bets on a betting exchange can be viewed as a system in which {\em opinion dynamics} is a key consideration. There is a long-established literature on opinion dynamics, but it tends to involve simple abstract models typically involving some number of agents (the population) interacting in such a way that the opinion of Agent A1 may to some extent influence the opinion of Agent A2, depending on specific circumstances, and possibly A2 would at the same time exert some influence on the opinion of A1, and as the model evolves through time some sequence of these pairwise opinion-altering interactions happens in the model system such that the overall distribution of opinions in the population changes in interesting ways over the duration of the simulation experiment. Much of this opinion dynamics work has been directed at identifying the conditions under which consensus can be reliably reached (e.g., \cite{degroot_1974}) and the conditions under which extreme opinions might form and spread within a population (e.g., \cite{deffuant_2006,meadows_cliff_2012,meadows_cliff_2013}). 

Opinion dynamics modelling techniques have only very recently been integrated into models of financial markets: see e.g.\/ \cite{lomas_cliff_2021} which was motivated by the recent groundbreaking work of Nobel-Prize-winner Robert \cite{shiller_2017,shiller_2019} on {\em Narrative Economics}, the attempt to better understand economically unfathomable behaviors (such as investing in an intrinsically valueless `asset' such as Bitcoin) as a function not of rational economic reasoning but rather as a function of the narratives, the stories, that economic agents tell each other and themselves in justifying their buying or selling decisions: these narratives are nothing more than external verbalizations of internally held opinions, so Shiller's work on narrative economics is manifestly entwined with opinion dynamics.  

Yet, to the best of my knowledge, the communities of researchers involved in agent-based modelling of opinion dynamics, and in the nascent field of narrative economics, have both thus far overlooked betting markets as a subject of study: this seems to me to be something of an oversight: surely a back or a lay is nothing more than an opinion expressed in the most concrete terms, a statement of how much money the bettor is willing to risk losing if their opinion is wrong? Surely a betting exchange, the market for an event, is just a set of monetised opinions, a bunch of bettors each willing to put their money where their mouth is. 

For this reason it seems reasonable to propose here that agent-based models of betting exchanges, such as BBE,  should be considered first-class instances of systems in which opinion dynamics can usefully be studied: all that is needed is for the population of simulated bettors to interact with one another in ways similar or identical to those already studied in the opinion dynamics literature: this would be a model of individual bettors interacting with one another, sharing tips on which competitor might win and which might lose, in a peer-to-peer fashion. That would be one aspect of BBE that closely matches traditional opinion dynamics studies, but the centralised order-book of the betting exchange, the market for a specific event, is a second form of opinion-influencer that is not traditionally studied in the opinion dynamics literature: surely if each bettor active in the market can view the array of backs and lays that make up that market as it changes over time, then it is entirely plausible that the instantaneous distribution of bets, or any patterns of changes in that distribution, could or should have an influence on the opinions of at least some bettors.  

To give an extreme example: bettor B1 may start out quite sure of her opinion that horse H1 will win the race, and she backs it accordingly; but if partway through the in-play betting on this race the amount backing H1 stays constant but the amount wagered in lays against H1 suddenly goes up sharply, then (human nature being what it is) there is a fair chance that B1's opinion of which horse is most likely to win will change, and she might hurriedly make a fresh bet to try to compensate for this. Here the distribution of stakes in the market is a reflection of the bettor-population's overall collective sentiment about the outcome of the event, and that representation of the sentiment in turn can alter the opinion of some or all bettors in the market: this feedback loop could be positive or negative for any one bettor, depending on that bettor's dispositional characteristics such as her certainty or confidence in her own opinion. Analysis of collective sentiment has been an active topic of research in financial markets for many years (see e.g.\ \cite{mitra_yu_2016,pozzi_etal_2016,birbeck_cliff_2018,liu_2020}), but sentiment analysis of bettor opinions from time-series of betting-exchange order-books seems to be a lot less researched.

Furthermore, with only the slightest change of language, we could say that when a bettor B1 backs some outcome ${\cal O}$, she is {\em predicting}\/ that ${\cal O}$ {\em will}\/ happen; and similarly that when a bettor B2 lays ${\cal O}$, B2 is predicting that ${\cal O}$ {\em will not} happen. In this characterization, there is no requirement for ${\cal O}$ to be a sporting event; it could as easily be the outcome of a political election, or which movie will win this year's best-picture Oscar award, or what the annual profits of a particular company will be when they are next reported, or what that company's share-price will be three months from now. This exposes the natural link between betting exchanges and online prediction markets such as the long-established \cite{iowa_2021}, and more recent entrants such as \cite{predictit_2021} and \cite{smarkets_2021}: for a review of research issues of prediction markets in corporate settings, see \cite{oleary_2012}. Technically, prediction markets usually allow the participants to trade in a {\em contingent security}, i.e.\ to buy or sell units of a tradeable asset that will have a specified nonzero value if the specified outcome does occur (that is, the end-payout is contingent on the outcome) and will have a zero value for all other outcomes, and hence any such prediction market could also be argued to represent a form of betting system. Moreover, the fact that the outcome ${\cal O}$ could be a financial outcome, concerning the price of an asset or the balance-sheet of a company, brings us back to the world of finance, in which a simple contingent security is more commonly known as a {\em binary option}\/ (binary because it pays either something or nothing, depending on some specific outcome condition): most exchange-traded derivative contracts such as futures and options are, put bluntly, just glorified gambling -- even if the participants in those markets like to think differently.

\section{Related Work}
\label{sec:litrev}

I have searched online  to the best of my ability for academic papers describing something similar to BBE, but have found no good matches. It is the case that there are various simulators offered by computer-gaming companies and by betting-exchange operators which allow a bettor to bet on simulated sporting events, using simulated money, either for pure fun or to practice betting without risking loss of real money, and several bookmakers also offer real-money betting on ``virtual" sporting events where the competitors are synthetic within a simulation or e-sports computer-game, but these such simulators do not offer the possibility of modelling each bettor's changing opinions and issuing of new bets during an in-play betting session, and are not research tools.

Synthetic horse-races have a long history. As reported by \cite{agini_2020}, an epidemic of foot-and-mouth disease (known in some countries as hoof-and-mouth) among farming livestock in the UK in 1967 prompted a suspension of all British horse-racing for several months, and the British Broadcasting Company (BBC) broadcast two computer-simulated horse-races, with live commentary: the {\em King George VI Chase}\/ and the {\em Massey Ferguson Gold Cup}, described in more detail by \cite{pitt_2020}: given the limited and expensive computer technology available in 1967, the simulation was extremely simple by today's standards.  Another UK foot-and-mouth outbreak in 2001 coincided with the development of virtual horse-races with sophisticated (in comparison to 1967) 3D computer-generated imagery: see e.g.\ \cite{bulkley_2003,men_2004}, and these proved so popular that they retained a loyal following of bettors even when live horse-racing had resumed, and by 2007 betting on virtual horses accounted for 17\% of all horse racing bets in the UK, with bookies' takings from bets on what punters fondly referred to as the ``plastic ponies'' rising to more than \pounds700M (\cite{bowers_2007}). Virtual sports are now routinely included as sports-betting events, e.g.\ Betfair offers betting on synthetic football/soccer matches, and on races involving synthetic horses, synthetic greyhounds, and synthetic racing motor-cars. As \cite{agini_2020} reports, live horse-racing was again suspended for the {\sc Sars-Covid-19} pandemic in 2020-21, and the UK's world-famous {\em Grand National}\/ 2020 horse-race was broadcast on primetime TV as a synthetic/virtual event. None of these synthetic horse-racing simulators are public-domain open-source research tools, and none were built to function as synthetic data generators.

BBE's simulation of a track-racing event is a deliberately minimal mathematical/algorithmic model, intended purely as bare-minimum mechanism capable of generating data that are plausibly similar to those that would be expected from a real track-racing event. It is described in detail in Section~\ref{sec:racesim}, but the race simulator is not in itself intended as a contribution to the simulation literature. As was mentioned in the previous paragraph, there are various simulators made available to bettors for pure fun, for practice-evaluation of betting strategies intended for subsequent deployment on real race-betting, and for real-money betting on computer-synthesized racing events presented to the bettors via CGI-rendered movies of the simulated races. To be of value to the user, the practice and for-fun systems involve some kind of simulation of betting on an exchange, at least from the user-interface perspective, whereas the real-money betting on synthetic CGI sporting events involve actual links to a real-world betting exchange, but these are all proprietary systems: their source-code is not open-sourced, and they are not offered as (and typically cannot be used as) generators of synthetic data for use in subsequent AI/ML-based research on automated betting strategies. Despite repeated and extensive online searches, I have found no reports of a sports-betting simulator that is even tangentially comparable to BBE. 

Despite this, there is some solace to be found in the literature on simulated financial exchanges: given the manifest similarity between betting exchanges and financial exchanges, the literature on financial-market simulation is of some relevance, and indeed in that field there is a tradition of researchers freely releasing public-domain open-source simulators as common platforms for research. Accurate simulators for the kind of LOB-based financial exchange explained above are found in the publications and associated open source-code releases for the simulators {\em OpEx}\/ (\cite{deluca_etal_2011_foresight,deluca_cliff_2011_ijcai,deluca_cliff_2011_icaart}), {\em BSE}\/ (\cite{cliff_2012_bse,cliff_2018_bse}), and {\em ExPo}\/ (e.g., \cite{stotter_etal_2014}), which each aimed to make public and free the kind of financial-exchange simulator that was previously private and proprietary when used in experiments conducted at major corporate research lab such as IBM (e.g., \cite{das_etal_2001}) and Hewlett-Packard (e.g., \cite{cliff_preist_2001}). The BBE betting-exchange simulator described in detail in the remainder of this paper has benefitted from the lessons learned from (and the open-source code releases of) OpEx, ExPo, and BSE. 

The similarities between financial markets and betting markets, and factors of interest that are common to the two types of markets, have been studied by academic economists, statisticians, pollsters, and forecasters -- see, for example: \cite{piron_smith_1995,hurley_mcdonough_1996,thorp_2006,camerer_2008,thompkins_etal_2008, palomino_etal_2009,franck_etal_2010,smith_vaughanwilliams_2010,haahr_2011,brown_etal_2016,flepp_etal_2017,sung_etal_2016,wall_etal_2017,hanke_etal_2018, restocchi_2018_phd} and \cite{elaad_etal_2020}. 

Although all of these studies involved analysis of data from real-world betting exchanges (or other gambling mechanisms) rather than working with synthetic data from a simulator. When such academic work does talk of simulation studies, this almost always involves the use of historical real-world time-series data as input to a process that then simulates a betting strategy operating on that real-world data over a number of days or weeks; I know of no such studies that simulate some number of heterogeneous bettors each with their own strategies for in-play betting. 

Possibly one reason for this apparent lack of agent-based models of betting-exchange systems is the lack of any literature on individual bettor-agent models or strategies. This is in stark contrast to the literature on autonomous trader-agents for financial markets where there is a 30-year tradition of publishing details of trader-agent strategies, each intended in one way or another as an improvement on the previous best. A tradition has emerged in this body of literature of giving each different strategy an abbreviated ticker-code style of name; the most notable strategies in the public domain are: Sniper (\cite{rust_etal_1992}); ZIC (\cite{gode_sunder_1993}); ZIP (\cite{cliff_1997_zip}); GD (\cite{gjerstad_dickhaut_1998}); RE (\cite{erev_roth_1998,pentapalli_2008}); MGD (\cite{tesauro_das_2001}); GDX (\cite{tesauro_bredin_2002}); HBL (\cite{gjerstad_2003}); AA (\cite{vytelingum_etal_2008}); SHVR and GVWY (\cite{cliff_2012_bse,cliff_2018_bse}); ASAD (\cite{stotter_etal_2013}), and PRZI (\cite{cliff_2021_przi}). All of these trader-agent models were essentially hand-designed, although from the mid-1990s onwards it became commonplace to incorporate adaptation mechanisms drawn from machine learning or statistical analytics, so that each individual trader could adapt to the particular market circumstances it found itself embedded in, and so that traders could respond to changes in market conditions as they happened. Much of this work on artificial trader-agents (now commonly referred to as {\em robot traders}) was intellectually underwritten by the large body of prior work in experimental economics, in which controlled repeatable empirical studies teased out the fine details of human trading behaviors in realistic laboratory models of financial markets (see, e.g.: \cite{smith_1962,smith_2000,kagel_roth_1997,plott_smith_2008}; and \cite{hommes_lebaron_2018}).

In comparison, for betting markets, a substantial body of research has been published in various fields that reports on empirical studies of the behavior of actual human bettors (see, e.g., \cite{kanto_etal_1992,swidler_shaw_1995,bradley_2002,jullien_salonie_2008,choi_hui_2014,feess_etal_2015,brown_yang_2016,suhonen_etal_2018} and \cite{merz_etal_2020}), where there is a common concept of the {\em representative bettor}, i.e.\ an idealisation of the betting behavior of the average bettor, and in which prospect theory (\cite{kahneman_tversky_1979,tversky_kahneman_1992}) has been a major influence. In this area of the literature, a fair amount of effort has been expended on exploring and explaining the {\em favorite-longshot bias}, where bettors tend to undervalue favourites (outcomes that have short odds, high probability) and overvalue outsiders (outcomes that have long odds, or low probability), a bias that is seen repeatedly in betting markets and in financial markets, but there is not a long-established literature that is comparable to the body of work on robot traders for financial markets, where researchers developed the succession of specific automated trading strategies listed in the preceding paragraph. 

For many years, researchers interested in the microstructural dynamics of financial markets (e.g.\ \cite{ohara_1995,dejong_rindi_2009}) have analysed high-resolution financial-exchange data and have identified a number of statistical characteristics that are commonly referred to as the ``stylized facts'' of financial market data (e.g.\ \cite{terasvirta_zhao_2010}) such as high kurtosis or slowly decaying autocorrelation. Any synthetic data generator for financial markets would (and should) be judged at least partially on the extent to which it can generate data series that exhibit the same stylised facts.

Unfortunately, as far as I have been able to determine, there is no body of research that identifies similar stylized facts in the time-series data from in-play betting on track-racing events. The closest I have found is the recent PhD thesis by \cite{restocchi_2018_phd}, who analysed data from prediction markets for political events. As was explained above, prediction markets are close relatives of in-play betting markets, but the two are typically not identical because the opportunity sets available to market participants differ. \cite{restocchi_2018_phd} did find some statistical characteristics in the political prediction market data that bear a reasonable comparison to the stylized facts of financial markets, but the extent to which in-play betting markets exhibit stylized facts, and the nature of those facts if they do exist, seem currently to be unknown: this is a point returned to in Section~\ref{sec:furtherwork}.

In recent years there has been a growing body of research publications exploring the use of statistical approaches, machine learning, and/or artificial intelligence, in betting markets. Various authors have reported mathematical or algorithmic approaches to profitable betting or trading on betting exchanges, often involving machine learning; examples include:
\cite{ioulianou_etal_2011, 
aruajosantos_2014,
bunker_susnjak_2019,
hubacek_etal_2019,
axen_cortis_2020, 
goncalves_etal_2020,
hubacek_sir_2020,
wheatcroft_2020}; and 
\cite{wilkens_2020}. However all of these studies work from databases of historical odds time-series from one or more betting exchanges, and none of them report on methods for in-play betting: in this sense, they are comparable to automated methods for identifying buy and sell signals from analysis of historic time-series of daily (as opposed to intra-day) price-movements in financial markets. 
Such an approach is perfectly valid, of course, but it gives little or no insight on how best to trade second-by-second in a fast-moving situation such as a market for a heavily-traded asset, or an in-play betting market for an event that is underway. Furthermore, such an approach fails to capture the close-coupled feedback loop where events occurring mid-race cause some bettors to shift their positions, placing new in-play bets, which are then visible to other bettors in the market, causing those other bettors to also adjust their positions. 

In comparison, the number of authors who have reported research on in-play betting is small. \cite{easton_uylangco_2009} present an analysis of in-play betting on tennis matches, but concentrate on comparing the time-series of in-play betting-exchange odds to the predictions made by a statistical model previously described by \cite{klaassen_magnus_2003}, thereby validating that model. The comprehensive PhD thesis by \cite{tsimpras_2015} presents a betting optimization system called {\sc Sportsbet}, and an associated domain-specific programming language called {\sc Ubel}, which are demonstrated in use for in-play football and tennis matches, and also for ex-ante betting on horse-races, working on both historical and real-time data fed from Betfair: there seems no reason to doubt that {\sc Sportsbet} could be used for refining betting strategies for in-play betting on race events, but none are reported by \cite{tsimpras_2015}; unlike BBE it cannot act as a synthetic data generator (SDG) because it was not designed to do so, and it is not available as an open-source research tool. Similarly, the system described by \cite{dzalbs_kalganova_2018} seems to have the capability to operate on in-play data, but the strategies that are developed in that paper are all dedicated on placing bets before the start of the race; and, like {\sc Sportsbet}, it is neither an SDG nor an open-source research tool. 

Fundamentally, even the highest-resolution time-series of in-play betting prices for a specific event is only half the story: without a similarly accurate record of how the event itself played out (e.g., second-by-second records of the positions of the competitors on the track), there is simply nothing to correlate the betting activity against. In principle, such data could be generated automatically via an appropriate computer vision system: videos of real races could be analysed to generate time-series of competitor-positions over time, either absolute distances along the race-track, or rank-ordered positional data (i.e., who is in first place, who is in second, and so on), but no such competitor time-series data seems to be available in the public domain. In the absence of such data, there is a need for a simulator that (re-)creates the kind of events that bettors like to bet on. Although in principle we could later extend BBE to incorporate simulations of sporting events such as soccer games or tennis matches, we started with the track-race events because they are relatively straightforward to characterize mathematically, as described in the next section.

\section{The Race Simulator}
\label{sec:racesim}

As currently configured, BBE is an abstract minimal model of some number of bettors interacting via a betting exchange to back and lay bets on the outcomes of racing events. The model is sufficiently minimal and abstract that in principle it could be interpreted as a model of gambling on horse-races or greyhound-racing, two sports in which betting is deeply embedded; or it could equally be interpreted as a model of gambling on races between vehicles such as in Formula-1 or NASCAR car-racing, motorbike racing, cycling races; or human track-and-field athletics running-races; or any other type of event where some number of participants are started at the same time and then compete to cross a finish-line first. There is nothing in our model that specifically limits us to one specific type of race, so we will talk here just of races and competitors. 

Let $\cal C$ be the pool of competitors, and let the number of competitors in $\cal C$ be $C=|{\cal C}|$. For any one race, denoted by subscript $r$, some number $n_r$ of competitors are drawn from ${\cal C}$ and then that subset of competitors, denoted ${\cal C}_r$,  compete by racing along a one-dimensional track of specified length $L_r$: the position along the track  of competitor $c$ at time $t$ is a real-valued distance denoted by $d_c(t)$, and the state of the race at time $t$ can be summarised by the vector $\vec{d}(t)$ in which the $i$th element is $d_i(t)$ and hence $|d(t)|=n_r$. Individual competitors are merely represented as points along the track: they have no physical extent in our model, although they can impede or block one another's progress, as described further below. 

A competitive race starts at time $t=0$, and the clock then continues to run until the last-placed competitor $c$ achieves a position $d_c(t)>= L_r$ -- i.e., the race ends when the slowest competitor crosses the finish line; in-play betting may be specified to end at that at time, or possibly when an earlier condition is met, such as the third-placed competitor passing the line. The initial value of $\vec{d}(t)$ might be a zero vector (i.e., all competitors lined up on the start-line, or all horses stood in a starting-gate) or not (e.g. some motor-races start with competitors spread out spatially over a ``grid'' with the better competitors to the front; and some horse-races start with the competitors gathered stationary behind a tape ``barrier'', at various distances from the tape). 

Each competitor's progress within a race is governed by a discrete-time process such that $d_c(t+\delta_t)= d_c(t)+S_c(t)$ where $S_c(t)$ is a function that generates a forward-step for competitor $c$ at time $t$: $S_c(t)>0$ at all times, to ensure that the race will eventually end, and should usually be a stochastic function so that the outcome of the race cannot be determined precisely at the start. For example, using ${\mathbb U}(lo,hi)$ to represent a uniformly distributed random variable over the range $[lo, hi] \in {\mathbb R}$, competitor $c1$ might have $S_{c1}(t)={\mathbb U}(10,20)$ while competitor $c2$ might instead have $S_{c2}(t)={\mathbb U}(1,25)$: given the specifications of these two $S$ functions, we can say that one competitor is more or less likely to cross the finish line first on the average, but we cannot say for sure who will win a specific individual race. 

In real races, any one competitor will be more or less suited to the details of a specific race: some competitors will do better in shorter races, while others will prefer longer distances; some will prefer flat courses, others will prefer undulations or hills; some will prefer racing in warm weather, others would prefer it to be cool; and so on. For studies of distance and pace preferences in horse-racing, see \cite{bolton_chapman_1986,benter_etal_1996,edelman_2008,edelman_2009}, and for analysis of how racetrack surface conditions affect racehorse performance see \cite{maeda_etal_2012}. To model these effects, BBE allows some number $f$ of performance-factors that modulate the $S_c(t)$ function for a competitor: in any one race $r$, the $f$ factors will each take on specific values (i.e., a particular length, a particular flatness, a particular temperature, etc) but these are all abstracted and normalized to simply be a vector $\vec{f}_r$ of length $f$, where each value in the vector lies in the range $[0,1]$, and each competitor's $S$ function is extended to incorporate $\vec{f}_r$, s.t. $d_c(t+\delta_t)=d_c(t)+S_c(t,\vec{f}_r)$. One simple illustrative instantiation of how $f$ can be used within the $S$ function is for each competitor $c$ to have a $f$-dimensional {\em preference vector} $\vec{p}_c$ specified when the competitor is created, and for the step-size to be modulated by the Cartesian distance between $\vec{p}_c$ and $\vec{f}_r$, e.g. $S_c(t,\vec{f}_r)=(k-|\vec{f}_r-\vec{p}_c|)\cdot {\mathbb U}(d_\text{min},d_\text{max})$ with the constant $k>0$ -- i.e., the farther the current race's factor-vector is away from the individual competitor's preference vector, the smaller that competitor's step-sizes will be in this race. 

More generally, we can say that each competitor $c$ has a {\em preference function}\/ ${\cal P}_c(\vec{f}_r, \vec{p}_c) \mapsto [0,1] \in {\mathbb R} $ which gives a multiplicative coefficient that can reduce the stochastic step-size taken by that competitor on each timestep by an amount that depends on the degree of mismatch between $\vec{f}_r$ and $\vec{p}_c$; and we can say that the stochastic step-generating function is denoted more generally as $\delta(\vec{v}_c)$ where $\vec{v}_c$ is $c$'s vector of parameters for whatever distribution is used for the step-generator, e.g.\ in the example given above, we'd say $\vec{v}_{c1}=[10, 20]^T$ and $\delta() = {\mathbb U}()$. 

For the races to present interesting betting opportunities we need the abilities of the various competitors to be sufficiently closely matched (given the race conditions) that the outcome is not obvious a priori. In horse-racing, jockeys wear weights as handicaps, and many races are restricted to particular types of horses (e.g.\ ``female horses no older than three years who have not won more than \$500 in the past''). These measures are intended to approximately balance the field, so that it is not absolutely obvious at the outset which runner will win. Our simulations are similarly constrained, otherwise we couldn't realistically expect our simulated bettors to interact with one another very much. 

Because we are interested in simulating in-play betting, we also add a within-race dynamic to each competitor's performance. This allows us to model, for example, situations in which a competitor takes some time to build up speed to maximum pace, and/or grows tired toward the end of the race (after an initial plateau, performance starts to gradually reduce) and/or makes a final burst of effort when in sight of the finish line (i.e., performance is briefly increased, until the end of the race). We model this by introducing for each competitor a {\em responsiveness} function ${\cal R}_c(t)$ which modulates ${\cal P}_c$. As just described, responsiveness might vary endogenously (i.e., due to factors internal to the competitor) but we are also interested in situations where responsiveness is affected by exogenous factors, i.e.\ due to causes external to the competitor: in the extreme, an exogenous factor could reduce a competitor's  responsiveness to zero, for example if a horse refuses or falls at a fence in jump-racing, or if one race-car is crashed into by another competitor at a tricky corner. For this reason our responsiveness function ${\cal R}_c$ for competitor $c$ should also take into account not only $c$'s current position but also the positions of all other competitors in the race, i.e.\ the full vector $\vec{d}(t)$. Then ${\cal R}_c(t,\vec{d}) \to {\mathbb R}^+$ gives a mechanism for modelling situations in which one competitor might start at a fast pace but slow toward the end of the race, while another might start slow and speed up at the end.

As described thus far, our simulated races are similar to short-distance sprints on an athletics track, where each runner is assigned their own lane to stay in, and hence there is an expectation that there will be no physical interactions between the $n_r$ competitors in the race: and, given the lack of any interaction between the competitors, the winner could in principle be decided by a sequence of $n_r$ single-competitor time-trials. However in longer-distance human athletics races, and in pretty-much all animal or vehicle races, the runners are not constrained to individual lanes and hence they can interact with each other on the track to create unexpected outcomes: the favourite at the start of the race might find themselves boxed-in by other competitors (this routinely happens in animal racing and in vehicle-racing too), or delayed trying to find an opportunity to overtake/lap a slower competitor (common in car races), and so a fast competitor may be slowed by one or more slower competitors immediately in front of it: I'll refer to that as {\em boxing in}. Similarly, another competitor coming up from the rear, threatening to overtake, might spur a burst of speed from the competitor in danger of being overtaken: I'll refer to this as {\em spurring on}. 

Our model of in-play betting would be made much richer by introducing some kind of in-race competitor interactions on the track, modelling boxing-in and spurring-on, so let's do that. 

Let $P1$ denote a competitor that is leading a race, $P2$ denote the competitor that is currently in second-place, and more generally let $Pn$ denote the $n^\text{th}$-place competitor. Then to model obstruction, boxing-in, and spurring-on, we can specify the $S_c$ function such that a competitor's forward-step size is determined at least in part by the proximity of other nearby competitors in front of and behind that competitor. 

Considering first the issue of obstruction, a $P1$ competitor's progress isn't interfered with at all; a competitor $P2$ with ``clean air'' in front of it (i.e., a long distance behind $P1$) is similarly unhindered; but a $P2$ immediately behind the $P1$ (or, more generally, any runner in position $Pn$ that has one or more runners at positions $P(n-1)$ or less in front of it and nearby) has a much higher chance of being impeded. This needs to be probabilistic (it's not {\em guaranteed}\/ that a competitor will always be impeded by another runner in front) and should ideally take into account the number of runners in close proximity, as well as their distances. For that reason, and also because some types of race require the competitors to each start in a designated lane/position and possibly also to remain in-lane until some condition is met (e.g. in Olympic 800m track-running, the competitors each start in individually-assigned lanes and must remain in them until they exit the first turn of the track), the function that models obstruction will also need the current track-positions of all the competitors, $\vec{d}(t)$: for modelling obstruction, we'll refer to the nearest competitor individual in front as $i^{+}$.

To model spurring-on, we simply need each competitor $c$'s responsiveness function ${\cal R}_c$ to be sensitive to the distance to the nearest competitor behind, denoted by $i^-$: if $i^-$ gets too close, less than some threshold distance $\theta_{c-}$, then $c$ will increase its value of ${\cal R}_c$, speeding up in an attempt to out-run $i^-$. Again, this should be nondeterministic: $c$ might forever out-run $i^-$, or it might not. 

Equation~\ref{eq:S} illustrates how the pace of the nearest competitor in front, $i^+$, can limit the forward step size if that competitor is too close.

\begin{equation}
S_c(t,\vec{f}_r,\vec{d}) = 
\begin{cases}
	{\cal R}_c(t,\vec{d}).{\cal P}_c(\vec{f}_r,\vec{p}_c).{\delta}(\vec{v}_c) & \text{if } \Delta_c(t)>\theta_{c+} \\
	{\cal R}_c(t,\vec{d}).\delta_{\text{min}}   & \text{otherwise.}
\end{cases}
\label{eq:S}
\end{equation}  
Where $\theta_{c+}$ is $c$'s threshold distance for being delayed by a slower-running competitor in front of it (i.e., if the distance to $i^+$ is more than $\theta_{c+}$ then $c$ is not delayed by that competitor); and $\Delta_c(t)=d_{i^+}(t)-d_c(t)$ is the distance to the nearest competitor ${i^+}\neq c$ who is in front of $c$, i.e.: 
\[
i^+ = \underset{i \in {\cal C}_r}{\arg\min}(\Delta_c(t), \forall i : d_i(t)>d_c(t)),
\] 
and 
\[
\delta_{\text{min}} = \min (S_c(t-\delta_t,\vec{f}_r,\vec{d}),S_{i^+}(t-\delta_t,\vec{f}_r,\vec{d})),
\] which means that if $i^+$ is too close in front of $c$, then $c$'s step-size becomes limited by $i^+$'s step size, if $i^+$ is slower than $c$.

Illustrative results are presented in Section~\ref{sec:results}, where further details are given of the ${\cal R}_c$ function used to generate those results. 

While it would be desirable to derive clean closed-form equations for the probability of any specific competitor winning any particular race, this is simply not practicable for cases of genuine interest because the competitors are deliberately closely matched and hence are very likely to interact in nonlinear ways within any one race. Instead, we can empirically determine reliable estimates of the true underlying probabilities by running a sufficiently large number $N$ of repeated independent and identically distributed (i.i.d.) ``practice'' simulations of a specific race with a specific set of competitors, and then calculating summary statistics from the data thus generated: in the simplest case, if after $N$ i.i.d.\ practice simulations of the same race a competitor $c$ has won the race in $w$ of those simulations, then our estimate of the probability of $c$ winning this race in the one ``competitive'' simulation in which actual betting occurs is $w/N$, and this estimate could be used by a bettor to set the starting price (i.e.\ the {\em ex ante}\/ odds) for that competitor. Performing the necessary number of i.i.d.\ practice simulations is computationally expensive on a single CPU -- it will run slow, potentially slower than real time, but it's embarrassingly parallelisable so it could be done faster by spreading the work over multiple machines or on many-core processors such as is explored in \cite{cliff_etal_2021_emss} and \cite{hawkins_2021_MEng}. Alternatively, a single ensemble of $N$ i.i.d.\ practice simulations could be run to establish the ``ground truth'' for the race at this point in time, and then each bettor's estimate of the outcome could be formed by injecting bettor-specific noise and perturbation factors to distort the bettor's estimate away from the ground truth. If the noise and perturbation factors are heterogeneously distributed across the population of bettors, this would have broadly the same effect in the simulator for less computational cost. 

Just to be clear, there is no claim being made here that real human bettors run multiple i.i.d.\ simulations of a race in their head before they decide on the likelihood of a particular competitor winning a specific race, nor that they have the absolute ground truth magically revealed to them, which they then screw up: humans' evaluation of probabilities and risk, and the way in which humans form and change their opinions about possible future events, has been the subject of much study and the mechanisms described here are definitely not intended as a contribution to those studies. Rather, what is described here is workably simple algorithmic method for arriving at the same situation as we see in a betting market: i.e., some number of bettors each with their own privately-held internal opinion about the likely winner of a race.   

Manifestly, much of the dynamics of a betting exchange is driven by differences of opinion among the bettors. So long as there is sufficient inherent uncertainty in the outcome of a simulated race, the differences in opinion can be engendered in various ways: one method would be to give each bettor $b$ only a small number $N_b$ of i.i.d.\ practice-runs of the race in which to form its opinion of the odds of success for each competitor -- i.e.\ for the individual sample-size to be sufficiently small that the bettor's estimates vary significantly; another method would be for each bettor $b$ to maintain their own private internal estimates of the values of the various constituents of the function $S_c$ (such as the preference vector $p_c$) for each competitor in the race, where $b$'s estimates differ from the true values, such that when $b$'s $N_b$ i.i.d.\ practice-runs generate its estimates of the probability of winning for each of the competitors, the resultant probabilities differ from the true underlying probabilities because of $b$'s inaccurate estimates -- this would model a bettor who is consistently over-optimistic or consistently over-pessimistic about a particular competitor's chances in a particular type of race.

Of course, for in-play betting, the bettors will need also to revise their estimates of the probable outcome of the race for each competitor while the race is underway. In principle, this could be done on every time-step, although in practice it can save computational effort if each bettor revises its opinions less frequently than that. This can be done using the same method as was used to create the bettors' {\em ex ante} opinions about likely outcomes: each bettor runs $N_b$ practice-races to form an estimate of the probabilities of winning for each runner, given where they currently are on the track -- so if a competitor $c_1$ that started as the favourite (i.e., most likely to win) happens to have been delayed by other runners and is now in the mid-field, and the second-favourite competitor $c_2$ has pulled into a clear lead, the odds for $c_1$ winning should be lengthened and the odds for $c_2$ winning should be shortened. Each bettor compares its revised estimates of the odds with those that are on the betting exchange and maybe revises/cancels/adds bets to the betting-exchange order-book accordingly.

As described thus far, we have a way of creating some number of races where each race involves some number of competitors from ${\cal C}$: this is enough to generate a sequence of sporting events on which bettors can gamble, but we need also to have a similarly minimal abstract model of a population of bettors ${\cal B}$ who gamble on these events. Just as not every competitor in $\cal C$ will be involved in each race, so there is no need for every bettor to gamble on each event: for any one event, some number of bettors can be drawn from $\cal B$. Fundamentally, a betting exchange operates by bringing together bettors with opposing views or beliefs: if bettor $b_1$ believes that competitor $c_1$ will win the current race, and bettor $b_2$ believes that $c_1$ will not win the race, then $b_1$ would back (i.e., bet-to-win) $c_1$ while $b_2$ would lay (i.e., bet-to-lose) $c_1$. The exchange can then match $b_1$'s back with $b_2$'s lay, facilitate the payment from one bettor to the other after the outcome of the race is known,  and take a small commission for doing so: both $b_1$ and $b_2$ have found a willing counterparty to take the other side of their bet, so they are each happy customers (at least, until one of them loses).

In our simulation, each bettor $b_i \in {\cal B}$ makes predictions about the outcome of a race and bets on the basis of those predictions. Intuitively, the accuracy of an individual bettor's predictions can be situated on a continuum from making equiprobable random choices over the space of possible  outcomes for a particular race (thereby totally ignoring all available information about the nature of the race and about each of the competitors) through to a god-like omniscient bettor who has perfect information on all factors that contribute to the outcome of the race (which, in our abstract model, would be full and precise details of the race's factor vector $\vec{f}$, each competitor's preference vector $\vec{p}$, and the details of each competitor $c$'s step-function $S_c$). One way of distributing the population of bettors along this continuum is to initially make each bettor form equiprobable estimates of the likelihood of each outcome for a race, and then to randomly allocate each bettor some number $d$ of ``dry-run'' trials: in any one dry-run, the race is simulated and that bettor use the outcome of that simulated race to revise its estimate of the outcome when the race actually takes place. A bettor with $d=0$ remains a purely random bettor; a bettor with $d=1$ has one trial's worth of data to go on, which is better than nothing but is not as good as $d=10$ or $d=100$; in the limit, as $d$ approaches infinity, the trial-outcome information that is available to an individual bettor is so extensive that accurate estimates of the probability of each possible outcome for the race can readily be made. 

Each bettor is endowed with some initial funds to be used in making bets, and bettors then submit back or lay bets to the exchange for each race, and the exchange's internal matching engine duly arranges the array of orders received, matching backs and lays as appropriate, and after the race is over the exchange takes care of the necessary book-keeping, updating the balances of the various bettors to account for their wins and losses.  

As explained thus far, the simulator can be seen to consist of the betting exchange, the race simulator with its pool of competitors and races, and the population of bettors with their varying degrees of accuracy of estimating the likelihood of different outcomes. 

In comparison to a financial market, the uncertain evolution of the positions of the competitors in the race as it unfolds is analogous to the moment-by-moment movements in the price of some tradeable financial asset; and the internal mechanisms of the betting exchange implement functions largely identical to those performed by the matching engine in a financial exchange: i.e., receiving orders from traders and aggregating them at different price-points and anonymizing them and then arranging them into a public/published limit order book (LOB) is much the same as receiving bets from bettors and aggregating, anonymizing, and arranging them onto the ``order book" showing the ``market" for bets on a particular race. 

Thus far I've described the component within BBE that is analogous to the dynamically-varying asset-prices in a financial market. Next, I'll look briefly at the model of the central exchange platform that collects orders and aggregates them into the order-book; and after that Section~\ref{sec:bettors} discusses the largely uncharted waters of modelling bettors for in-play markets.  

\section{The Betting Exchange}
\label{sec:exchange}

Algorithmically, the betting exchange's matching-engine itself is largely a matter of ensuring accurate book-keeping. For each competitor in a particular race, a record is kept of all the back-bets and all the lay bets on that competitor that have been received by the exchange: the internal record of each bet includes the arrival-time of the bet at the exchange, the identity of the bettor, and the amount wagered. Once a bet is received and recorded at the exchange, and matched with a counterparty bet, it cannot be cancelled, but unmatched bets can be cancelled, and any bettor can submit more than one bet into the market for a particular event. When a race event starts and turns in-play, all unmatched bets on that event are cancelled.

The arrival-time of a bet matters because multiple bets at the same odds are processed in time-priority order. The matching process pairs up buys and lays: when a lay at a particular price arrives, it is matched with the oldest unmatched back at that price; and when a back arrives it is matched with the oldest unmatched lay at the same price. A bet for a large stake can be fully matched against multiple bets of the opposite direction for smaller stakes: e.g. a \$100 back might be matched with three lays at prices of \$30, \$50, and \$20, and the backer is then treated internally as having three separate bets on the exchange. Partial matches remain active, so if the \$100 back is initially only matched with lays of \$30 and \$50, the remaining \$20 back is held at the exchange awaiting the subsequent arrival of a matching lay. All bets that are unmatched at the close of betting expire and the stakes are returned to the bettors.

The ``market" for any one competitor $c$ in a specific race is formed by aggregating across all back bets of the same odds, and across all lay bets of the same odds, to calculate the total amount of unmatched money wagered on each back or lay at each specific odds, and that is then displayed as the set of odds and total unmatched stakes for $c$, in $c$'s row of the overall ``market" race-grid or competitor-table for that particular race, as was illustrated in Figures~\ref{fig:betlob} and~\ref{fig:betladder}: the ladder is the order-book for $c$, and the market grid for the race involving $N$ competitors is a set of $N$ such order-books.
 
When the event ends, stakes are collected from the accounts of bettors who lost their bets, and the funds thereby accumulated are then distributed to those bettors who placed winning bets. Because the exchange is a platform, taking neither side of the bet, the exchange makes its money by charging a small percentage commission fee (e.g.\ 5\%) on winnings; losing bets are not charged.

The BBE matching-engine is the most straightforward component: the functionality of a real-world betting exchange is well documented; and so there is relatively little latitude or room for creativity in the implementation of this component of the BBE simulator, which is why it does not take up much of the discussion here.  Indeed, it could be plausibly argued that the BBE exchange module is not a simulation of a betting exchange; it {\em is} a betting exchange (that is, the core matching engine in BBE is not an {\em abstraction} of the real thing, but is instead an {\em instance} of the real thing). Nevertheless there is an awful lot more lee-way, and untrodden turf, when it comes to modelling the other major component in BBE, the bettors, as discussed in the next section. 

\section{Modelling Bettors}
\label{sec:bettors}

As was noted in Section~\ref{sec:litrev}, in the research literature on agent-based modelling of financial markets there is a long history, stretching back over multiple decades, of researchers publishing details of new algorithmic trader-agent strategies that in one way or another are intended as improvements on prior work, and for which evaluation and comparison with earlier strategies has been relatively straightforward. In contrast, the absence thus far of public-domain betting-exchange simulators such as BBE has meant that there is no common platform for comparison or evaluation of in-play betting strategies deployed by bettor-agents within the simulation model; and this is likely to have contributed to the current situation in which there is no comparable history of research publications describing a succession of in-play automated betting strategies that each improve on past work. This is both a blessing and a curse: a blessing in the sense that there is lots to explore, lots of opportunities for interesting research and innovation; but a curse in that here we are pretty-much starting from scratch. 

For a bettor placing a back bet in a sports betting exchange's win-bet market for a horse-race, there are two fundamental questions that need to be answered: what horse do I think will win; and what odds should I back that horse at? For a bettor in the same market looking to place a lay bet, two very similar questions immediately present themselves: what horse do I think least likely to win, and what odds should I offer to lay that horse at? Section~\ref{sec:choose_horse} addresses the issue of selecting a horse; and Section~\ref{sec:choose_odds} discusses how, once a BBE agent has chosen a horse to bet on, it determines what odds to bet at; before that, Section~\ref{sec:bettor_basics} discusses the basic question of how many bettors we need for a liquid market.

\subsection{How many bettors?}
\label{sec:bettor_basics}
In setting up any agent-based model such as BBE, there is a fundamental question to be answered: how many agents will we need? This can be explored with a simple illustrative example. Let's refer to the betting-exchange's market for a race as {\em nonempty}\/ if each competitor has at least one back and lay matched for each competitor, i.e.\ if the order-book for the market contains no rows, no competitors, on which there is no bet currently in place. Consider a $N$-horse race with horses labelled as $H_1$, $H_2$, ... $H_N$. For a race's market order-book to be nonempty, we would need two bettors per horse, one to back it and one to lay it. In the simplest $N=2$ case, assuming each bettor can place only one bet, we would need bettors $B_1$ and $B_2$ to match their back and lay bets on $H_1$, and we'd need bettors $B_3$ and $B_4$ to match a back with a lay on $H_2$; and intuitively we can see that we'd need $2N$ bettors for an $N$-horse race to create a nonempty market if each bettor bets only once. If we allow each bettor to randomly choose both the horse it bets on, and whether to back or lay that horse, then it's simple to show that the probability $p$ of these random choices (assuming each choice is uniformly distributed) resulting in a nonempty market is given by $N!/(N^N)$. This gives $p=0.5$ for $N=2$ but $p$ rapidly approaches zero as $N$ increases, e.g. at $N=5, p=0.038$; and at $N=10, p\approx0.0004$; so for any race with a reasonable number $N$ of runners, it is obvious that the number of bettors $B$ we need for interesting market dynamics is $B>>N$. Indeed, referring back to Figure~\ref{fig:betlob}, if we want the row for a single competitor to have a minimum of $D$ distinct prices on both the back side and the lay side of the book (i.e., in Figure~\ref{fig:betlob}, $D=3$), then the minimum number of one-bet bettors we need is $2DN$, and that would give only a single bettor for each of the 2D prices for each of the N competitors ithat the betting market covers -- for those bettors to be matched with counterparties requires another $2DN$ bettors each taking the opposite view, and so a minimal lower-bound on the number of bettors is $B \geq 4DN$.

\subsection{Choosing who to bet on}
\label{sec:choose_horse}

Clearly BBE needs a population of bettor agents that are sufficiently diverse in their views, in the accuracy of their predictions, that the BBE market has decent liquidity -- i.e., that there is a good chance of a bettor finding a counterparty to take the opposite view and wager a stake. The discussion in Section~\ref{sec:racesim} explained how rational BBE bettors can use multiple simulations of the race to estimate odds before a race commences, and also (by simulating the race playing out from its current state at time $t$ until the race-end) as a way of attempting to estimate the likelihood of the remaining possible outcomes once the race is in progress, thereby enabling the bettors to take reasonable actions in their in-play betting. This approach, of running forward simulations from current conditions is an attempt at giving the BBE some sensible sense of {\em rational}\/ behavior. It seems fair to assume that in any reasonably well-populated betting-exchange market, a decent proportion of the bettors will be doing their best to act rationally (although, of course, some bettor's attempts at rationality will be better than others, the human mind's propensity for falling prey to various forms of bias being what it is). The ``representative bettor'' research discussed in Section~\ref{sec:litrev} provides us with cues, such as the favourite-longshot bias, for how to make BBE's bettor-agents plausibly imperfect. 

But in real-world betting exchanges, not every bettor is rational. Some will bet on a particular outcome because they like the name of a competitor; or because that competitor is carrying the bettor's ``lucky number'' as its identifier in the race; or because the bettor is a big fan of a particular competitor and they place bets paying out on success for that competitor as a token of loyalty, of fandom; or because the bettor chose what outcome to bet on by closing her eyes and sticking a pin into a list of competitors; and so on. It is clear then that BBE needs not only a model of bettor rationality, but also a decent model of bettor irrationality. Indeed, irrational bettors are extremely useful to the operators of a betting exchange, because they provide a supply of potential counterparties to take the opposite view to those held by rational bettors: luckily for the rational bettors, there is an irrational bettor born every minute. And of course, occasionally, an irrational bettor will strike it lucky. Once again, prior modelling in financial markets provides a comparison: many financial-market models will include some number of ``noise traders'', which act irrationally, alongside some number of traders who are intended to be acting more rationally; so a BBE simulation should include some number of ``noise bettors'', who make random and uninformed bets.
  
With noise bettors at one end of the spectrum, and with the multiple-simulation ensemble-modelling rational method of estimating the likelihoods of different outcomes at the other, we have a continuum of bettor sophistication. The latter class of bettor-agent, the most rational ones in BBE, have already been described above in Section~\ref{sec:racesim}: their estimate of the outcome of the race comes from aggregating the results from some number $d$ of i.i.d.\ ``dry-runs'' of the race simulator -- I refer to these as {\em Rational Predictors}\/ with $d$ dry-runs as RP($d$) bettors, where the higher the value of $d$ the more accurate the bettor's prediction of the outcome is expected to be.

While having bettors at both ends of this spectrum is better than only studying one end of the continuum, the dynamics of the betting exchange are likely to be richer if the population of bettors is spread out along this spectrum rather than all being concentrated at one or both ends. 

Although there are a decent number of research publications discussing the notion of the representative bettor, pretty-much all of that addresses issues of how humans bet before an event is underway. Unfortunately, as far as I have been able to determine, there is only very little research literature on human betting behavior during in-play events, and so any attempt at creating human-like in-play betting behavior is limited by the lack of available human-behavior data.

Nevertheless, there is literature in other fields that can usefully be drawn from:  the way in which track-racing has been abstracted in BSE, i.e.\ the modelling of the race as each competitor's progress along a one-dimensional number-line racetrack, means that in any one race each competitor's progress along the track can be treated as a time-series of distance measurements, and the introductory end of the vast literature on time-series analysis can then be mined for creating some minimal but plausible in-play betting strategies. One example is referred to within BBE as the ``Linex" strategy, because it involves {\sc lin}ear {\sc ex}trapolation: a Linex bettor estimates the current speed of each competitor at each timestep by taking the arithmetic mean of that competitor's step-size over the past $N$ seconds, and assumes that these speeds will each remain constant for the rest of the race; then, working from each competitor's current position on the track and the estimate of their current speed, the Linex bettor calculates a prediction of which competitor will cross the line first.
Linex is just one of several bettor strategies that I've coded into BBE, to get it started. Others, listed here in no particular order, include the following:

\begin{itemize}
\item
LW (Leader Wins): this bettor's view of the outcome of the race is that whichever competitor is currently in the lead will go on to win.

\item
UD (Underdog): this bettor predicts that the P2 competitor will win, so long as the distance to the race-leader is less than some threshold distance $D$: if the P2 racer falls behind by more than that, the Underdog bettor switches prediction to the P1 competitor. 

\item
BTF (Back The Favourite): this class of in-play bettor monitors the distribution of stakes in the market and predicts that the winning competitor will be the one that currently has the lowest odds, i.e.\ the market's favourite.

\item
RB (Representative Bettor): a bettor agent that is programmed to behave in ways consistent with research results on human betting behaviors: although there is comparatively little literature covering human behavior in in-play betting, there is \cite{brown_yang_2016} who note that, in the hurry and heat of the moment, humans tend to choose stakes that are nonuniformly distributed across the space of possible amounts but instead cluster on multiples of 2, 5, or 10; and the well-known {\em favorite-longshot} bias is also coded into BBE's RB Bettors. 

\item
ZI (Zero Intelligence): these bettors choose a competitor at random and assume it will win -- these are the ``noise traders'' of a betting market. 

\end{itemize}

Populating BBE with a suitably large number of these bettors, with randomly-varied values for their parameters (such as $D$ in {\em Underdog}, or $N$ in {\em Linex}) gives sufficiently rich variation in opinion while the race is underway that the dynamics of the in-play market is plausibly nontrivial.

\subsection{Choosing odds to bet at}
\label{sec:choose_odds}

The process by which two bettors, one a backer and the other a layer, interact and come to agree the odds for a bet can usefully be compared to the process by which a buyer and seller agree on a price for a transaction, whether bartering in a souk or issuing electronic orders in a model financial exchange. Say that the seller initially asks \$100 and the buyer initially bids \$50: both have signalled their intent, but both run the risk that some other trader might come in with a more competitive price (i.e., a lower ask-price or a higher bid-price) --- the mere threat of this might prompt either trader to quote again at a better price, in the hope of increasing the chances of finding a price that is acceptable to a counterparty trader, and so the usual sequence of events is that the prices of bids and asks will converge toward a final transaction price somewhere between the bounds established by the opening bid and ask. The convergence to a final agreed amount for the transaction is known as {\em price discovery}.

Similarly in a betting exchange lower-priced back-bets and higher-priced lays are more likely to find a match. A bettor $B_1$ might initially intend to back horse $H_1$ by sending a back-bet to the exchange with optimistically high odds, i.e.\ a high price $P_1$, and stake $S_1$, offering a large payout to $B_1$ if $H_1$ wins; but if similar bets already on the exchange do not get matched, and/or if other bettors come in with back bets at lower odds that are more likely to be matched by willing counterparties making the corresponding lay bets, then $B_1$ might revise her odds down, lowering the price to $P_2<P_1$, which would place $B_1$'s bet closer to the front of the book. A complementary process unfolds on the lay side of the market for $H_1$, where some bettor $B_2$ might initially lay at a low price, at low odds, and those low odds might fail to be matched, or might be improved upon by other bettors laying $H_1$ at a higher price, which gives an incentive for $B_2$ to raise her lay odds if she wishes to stay competitive in the market. 

Having now introduced all the components of BBE, let's look at some illustrative results.

\section{Illustrative Results}
\label{sec:results} 

\begin{figure}[p]
\begin{center}
\includegraphics[width=0.75\linewidth]{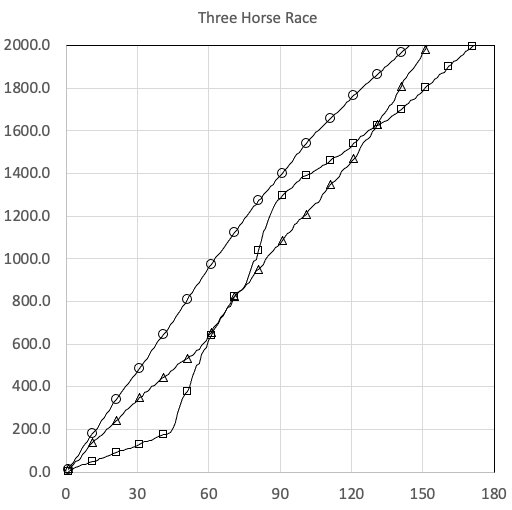}
\end{center}
\caption{Space-time plot for a three-horse race in BBE. The horizontal axis is time in seconds, and the vertical axis is distance travelled in metres: this race is over 2000m and lasts less than three minutes. The markers show progress every 10 seconds, and conveniently the name of each horse is also the shape of its marker. Here {\em Square}'s parameters have been set such that it gets off to a very slow start, but after 45s it picks up the pace and starts to gain rapidly on its nearest competitor, {\em Triangle}. At approximately t=60s {\em Square}\/ catches up with {\em Triangle}\/ but overtaking proves difficult, and for roughly 15 seconds {\em Square}'s progress is blocked by Triangle. Square then overtakes {\em Triangle} and rapidly closes in on {\em Circle}, who is in the lead. But {\em Square} fails to catch {\em Circle} and then goes off the pace, letting {\em Circle} pull away and with {\em Square} rapidly losing ground to {\em Triangle}, who pulls into second place just before the finish-line. Final result is {\em Circle} at P1, {\em Triangle} P2, and {\em Square} P3. The parameter values for this race were chosen to create an illustrative race with some dramatic turns of fortune occurring: this is not intended as a model of any real horse race.}        
\label{fig:race1plot}
\end{figure}

Figure~\ref{fig:race1plot} shows data generated by BBE's race-simulator, for a three-competitor race over 2000m, lasting around 2.5 minutes: the progress of the race is visualised as a plot of distance over time, and the interaction between competitors can clearly be seen. In this race the simulator parameters have been deliberately configured to give the competitor named {\em Square}\/ a manifestly eventful race: Square has a very slow start and then speeds up, but finds itself blocked and delayed by the two competitors in front, named {\em Circle}\/ and {\em Triangle}.  Conveniently, the competitors each have the same name as their marker on the graph.

\begin{figure}
\begin{center}
\includegraphics[width=0.75\linewidth]{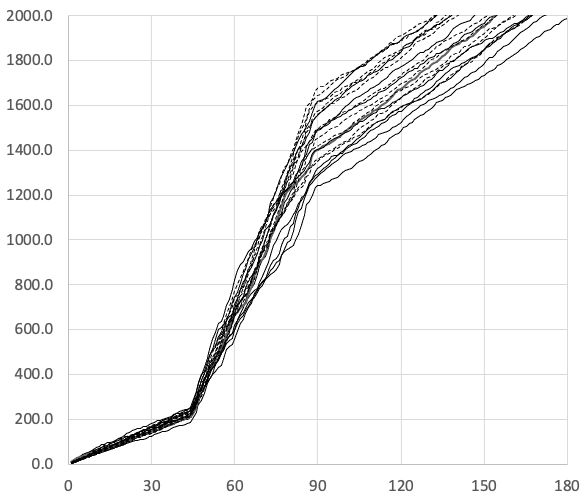}
\end{center}
\caption{ 
Multiple space-time trajectories for the horse called {\em Square} in i.i.d. repetitions of the three-horse race illustrated in Figure~\ref{fig:race1plot} -- the trajectories of the horses {\em Circle}\/ and {\em Triangle}\/ are not shown here. Axes are as for Figure~\ref{fig:race1plot}. Solid lines show {\em Square}'s trajectory in 10 races (one of which is the race illustrated in Figure~\ref{fig:race1plot}) in which {\em Square}\/ could lose ground and be delayed by slower-running competitors in front; dashed lines show 10 trajectories in which the race simulator was run with such ground-lost interactions disabled. In the absence of competitor interactions, the trajectories are more tightly clustered. Again, this is a deliberately extreme scenario in which {\em Square}'s 
responsiveness function 
${\cal R}(t,\vec{d})$ 
undergoes step-changes at $t=45s$ and $t=90s$, and is not intended as a simulation of a real race event.}
\label{fig:RACE1-tracks-interact}
\end{figure}

Figure~\ref{fig:RACE1-tracks-interact} illustrates the effect that interactions between competitors can have on the outcome of a race: it shows the space-time plot for competitor {\em Square} over 10 i.i.d. repetitions of the race illustrated in Figure~\ref{fig:race1plot}: the trajectories of the other two competitors are not shown, for clarity. This illustrates the degree of variation in outcome for {\em Square}, and the extent to which this variation is down to the ground lost due to interactions with other competitors.
 
\begin{figure}
\begin{center}
\includegraphics[width=0.75\linewidth]{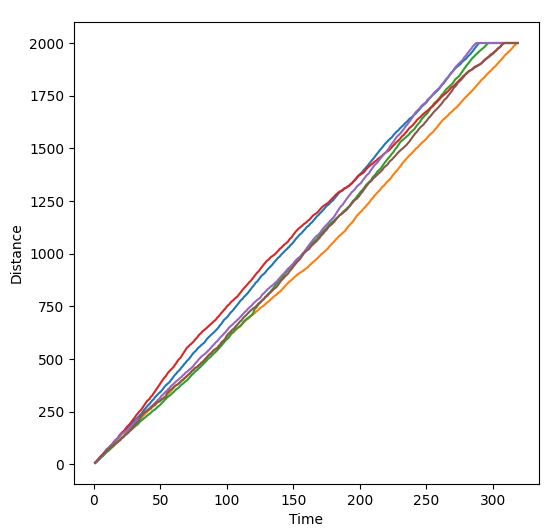}
\end{center}
\caption{Space-time plot for a more realistic, and apparently less dramatic race than that illustrated in Figure~\ref{fig:race1plot}. This is a six-horse-race, with each horse's space-time trace being shown in a different color, which again conveniently matches the name of the horse (so e.g.\ the horse called {\em Yellow} has a yellow trace in this graph). When each horse crosses the finish-line, it stops immediately, which is why in this projection each horse's space-time trajectory terminates with the horizontal straight line visible at distance=2000m. As can be seen, this style of plot wastes an awful lot of whitespace and obscures the fine-grained detail of how the race unfolds.}
\label{fig:6hr_raw}
\end{figure}

Although space-time plots such as Figures~\ref{fig:race1plot} and~\ref{fig:RACE1-tracks-interact} are a conventional and intuitive way of displaying a track-race, when the race is less dramatic (i.e., more realistic), they obscure a lot of important detail. To illustrate this, Figure~\ref{fig:6hr_raw} shows a conventional space-time plot for a more realistic race in which there are fewer severe changes in the speeds of the runners: all the significant/interesting action in this graph is heavily compressed along the diagonal, and there are large areas of uninformative whitespace. 

\begin{figure}
\begin{center}
\includegraphics[width=0.75\linewidth]{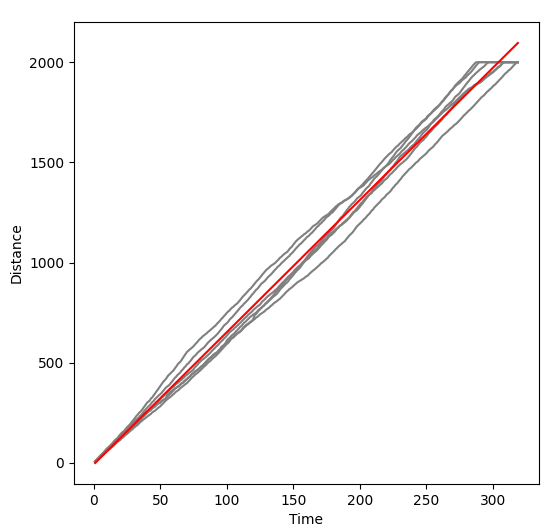}
\end{center}
\caption{The same six-horse race as was illustrated in Figure~\ref{fig:6hr_raw}, with each horse's individual space-time trace now plotted in the same shade of gray, and the least-squares linear regression line of best fit shown in red: this line can be thought of as a prediction of where a hypothetical representative constant-speed horse would be at each point in the race. This serves no practical predictive value, but it does give us a projection axis for re-visualising the actual horses' spacetime data, as shown in Figure~\ref{fig:6hr_rebased}.}
\label{fig:6hr_linreg}
\end{figure}

The same data can be shown in a different projection by treating each horse's trajectory as contributing a set of (time, distance) points to a scatter-plot of all the horses' space-time data for the entire race, and using linear regression on that set of space-time points to give a
`prediction' of where a constant-speed representative horse would be at each time-point in the race: this is illustrated in 
Figure~\ref{fig:6hr_linreg}. 

\begin{figure}
\begin{center}
\includegraphics[width=0.75\linewidth]{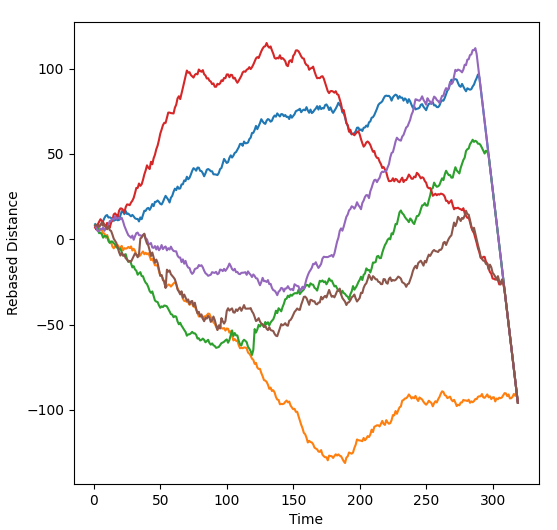}
\end{center}
\caption{The same six-horse race as was illustrated in Figure~\ref{fig:6hr_raw}, with the linear-regression `prediction' of the representative horse's position (as illustrated in Figure~\ref{fig:6hr_linreg}) subtracted from each horse's actual position: this rebased distance projection shows much more clearly the moment-by-moment changes in distance between the competitors, and their rank-order in the race at each timestep (the horse with the uppermost trace is in the lead, e.g.\ here the horse named {\em Red}\/ leads for the first 170 seconds or so, eventually being overtaken by {\em Blue}. When each horse crosses the finish-line, it stops immediately, which is why in this projection each horse's space-time trajectory terminates on the off-vertical straight line visible at times 280-320s. }
\label{fig:6hr_rebased}
\end{figure}

The linear-regression line is of no real interest as an actual predictor, but it does serve as a very useful axis along which the race spacetime data can be projected, as shown in Figure~\ref{fig:6hr_rebased}. This form of visualization makes much clearer the moment-by-moment changes in the inter-competitor distances, and also makes the rank-ordering (who is in the lead, who is in second-place, who is running last, etc) much clearer.

In the race illustrated in Figures~\ref{fig:6hr_raw} to~\ref{fig:6hr_rebased}, the responsiveness function ${\cal R}_c$ for each horse $c$ was set as follows: the total race distance was normalised to [0,1] and was partitioned into a set of phases for each competitor $c$ by randomly choosing the number of phase boundaries $p_c$ where $p_c = {\mathbb U}(2,4) \in {\mathbb Z}^+$ -- so some competitors might have two phase boundaries within the race (i.e., their responsiveness is in three phases over the course of the race) while others might have three or four, and then randomly choosing a percentage-race-elapsed for the boundary point of each phase. For any one competitor, its responsiveness in each of its phases is assigned at initialisation from ${\mathbb U}(0.7, 1.0) \in {\mathbb R}$, i.e., in each of its phases, each competitor has a baseline responsiveness of between 70\% and 100\%. This is intended to model the fact that some competitors might go off fast, surging at the start but then fading toward the end; while others might start slow and then put on a faster pace toward the end of the race. Each competitor's responsiveness phase boundaries, and the responsiveness levels within each phase, are fixed at initialisation, but at initialisation each competitor is also assigned a variation or noise parameter-value for both the phase-boundaries and the response-levels within each phase: this is used to introduce small degrees of variation of the actual phase boundaries and responsiveness levels on any one run of the race simulator, so that for example on three successive runs of the simulator a particular competitor may switch from a slow start to a faster pace at 36\%, 32\%, and 35\% of race-distance, and the new faster responsiveness might be 87\%, 85\%, and 81\% respectively.

\begin{figure}[h]
\begin{center}
\includegraphics[width=0.75\linewidth]{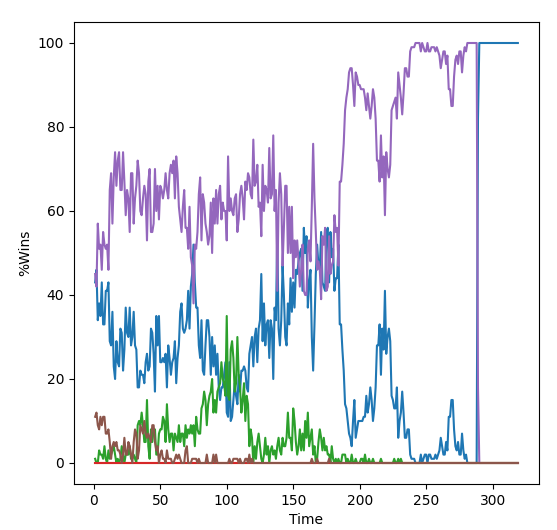}
\end{center}
\caption{Changes in the sentiment of an individual RP(d) bettor, with $d=100$ dry-runs per timestep, over the course of the race illustrated in Figures~\ref{fig:6hr_raw} to~\ref{fig:6hr_rebased}. The horizontal axis is time, and the vertical axis shows, for each horse, the bettor's currently estimate of the probability that the horse will be next to pass the finish-line, expressed as a percentage. This bettor's internal dry-run simulations revealed that {\em Purple}\/ was most likely to win throughout almost all of the race, despite its poor start (i.e., {\em Purple}\/ is a strong finisher, a ``closer'' in racing parlance).}
\label{fig:6hr-RPdBettor}
\end{figure}

Finally, Figure~\ref{fig:6hr-RPdBettor} shows a summary of the opinion of a single RP(d) bettor as the race in Figure~\ref{fig:6hr_raw} unfolds. 


\section{Further Work}
\label{sec:furtherwork}
 
This paper has focused on establishing the rationale for BBE, on surveying relevant prior literature, and on introducing the major components of the simulator. The brief illustrative sample output presented in the previous section demonstrates that we now have all the components in place for producing time-series of in-play betting markets at maximal temporal resolution (i.e., every substantive change in the state of the betting market can be captured in the data). In a companion paper to this one (\cite{cliff_etal_2021_emss}), I and my co-authors discuss three different implementations of BBE in depth, each of which is positioned at a different point along the continuum from being maximally accessible to non-expert programmers (and hence running comparatively slowly), to being engineered in such a way that detailed technical knowledge is required to alter and/or extend the code (but which runs two orders of magnitude faster than the maximally accessible version), and present comparative results. These three implementations are each being made freely available on GitHub, as public-domain resources for the research community: see \cite{cliff_etal_2021_emss,hawkins_2021_MEng,keen_2021_MEng} and \cite{lausoto_2021_MEng} for further details.

Having created multiple independent implementations of BSE, the most compelling avenue for further research is to conduct detailed statistical analysis of the dynamics of odds and stakes in in-play betting on horse-races. The aim of this research would be to establish a set of ``stylized facts'' for horse-racing markets that is comparable to the stylized facts, the statistical characterisation, of the dynamics of prices and volumes in financial markets. As was discussed in Section~\ref{sec:litrev}, as far as I can determine there seems to be no prior published research available that specifically deals with the stylized facts of in-play betting markets. Without a workable statistical characterization of the dynamics of real betting markets, it is impossible to calibrate BBE to real-world betting markets, and from this comes the risk that any system which uses BBE as a synthetic data generator (SDG) will adapt to features in the BBE data that are not present in the real world data, and/or will fail to adapt to features in the real world data that are not adequately reproduced in the BBE data. Because BBE is a {\em constructive}\/ SGD (i.e., the synthetic data is generated from causal mechanistic interactions in an agent-based model of the real-world system) rather than a {\em reproductive}\/ SGD (i.e., the synthetic data is generated from an opaque black-box method such as deep learning, having been trained on real-world data), it should be entirely possible to reason about any mismatches between the BBE data and the real-world data, and to adjust the BBE implementation to reduce the degree of discrepancy. Any simulation of a complex nonlinear system will typically go through a process of iterative refinement, a series of adjustments, before there is good alignment between the simulation and the system it models: BBE as reported here is at the start of that process, but there are reasonable grounds for being optimistic about its prospects. 

Once there is some measure of the extent to which BBE does capture the statistical nature of real-world in-play betting market data, and the degree of discrepancy between BBE data and real betting-exchange data has been reduced to within tolerable bounds, then BBE can be put to use for the work that it was built to enable: the use of AI and ML techniques for identifying profitable automated betting strategies. Work has already commenced on using BBE as the fitness evaluation function for an evolutionary computation approach to exploring the space of possible betting strategies: results from this will be reported in future papers. 

\begin{figure}[h]
\begin{center}
\includegraphics[width=0.95\linewidth]{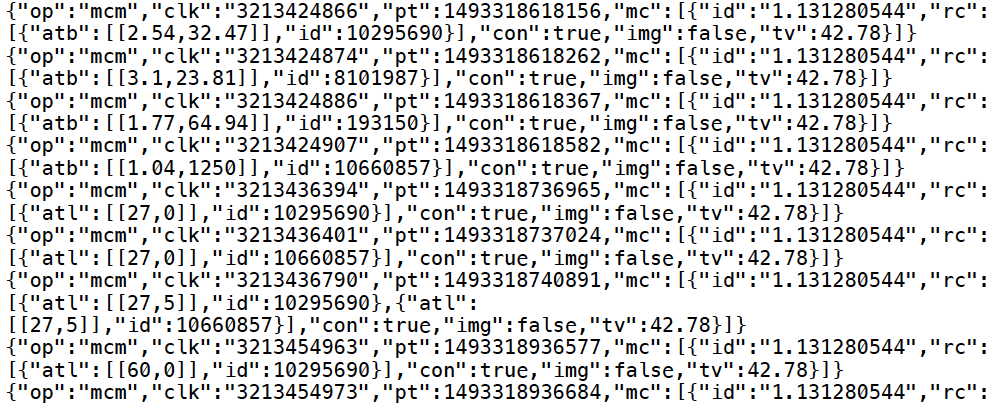}
\end{center}
\caption{A brief extract from one of Betfair's ``Pro'' (highest-temporal-frequency) data-files for a real horse race, available as a free sample from {\tt https://historicdata.betfair.com/\#/help}. This shows a sequence of eight {\em market change message} operations (abbreviated to {\tt "op":"mcm"}) which are changes to market prices, to competitors, or to the market definition. The {\tt "pt"} field is the {\em published time}, in milliseconds since the Unix epoch moment (00:00:00 on 1st January 1970). Betfair publish a full specification of their data-file format on their website, to which the reader is referred for further details: as is manifest in this extract, the data format is some way distant from being self-explanatory.
 }
\label{fig:betfairdata}
\end{figure}

Finally, one relatively straightforward direction for further work is to extend BBE so that the betting-exchange data-files that produces for each race are in the same proprietary format as used by one of the major commercial betting-exchange operators. Figure~\ref{fig:betfairdata} shows a short extract from one of the freely-available sample data-files made available by Betfair: as can be seen, the data is not exactly self-explanatory, but it does follow a straightforward published schema. If BBE data was produced in Betfair's native format then any software already written to analyse Betfair data could also be used to analyse BBE data, which could be an aid to the analysis of stylised facts. 

\section{Conclusions}
\label{sec:conclusion}

This paper has introduced BBE, an agent-based model simulating a group of bettors interacting via a betting exchange to make ``in play'' back and lay bets on the outcome of a race event, while that event is underway. To the best of my knowledge, BBE is the first simulator of its kind, in that no other in-race betting-exchange simulators are available as open-source public domain research tools, or synthetic data generators. BBE can be used to generate large-scale data-sets on sub-second temporal resolution from arbitrarily large number of simulated races: this enables the very low-cost generation of extremely large data-sets that can be used for training data-intensive machine learning systems in the search for profitable automated trading systems. Three independent implementations of BBE have been developed and are being released as open-source on GitHub, as documented in more detail by \cite{cliff_etal_2021_emss,hawkins_2021_MEng,keen_2021_MEng}; and \cite{lausoto_2021_MEng}. Future papers will report on the results from generating and using such BBE-generated data; and, with the BBE source-code available on GitHub, my hope is that other researchers will use BBE as a common platform, facilitating ready replication of results, and also that other researchers contribute to extending the BBE codebase as required.

\section*{Acknowledgements}
Thanks to Veyndan Stuart, who worked with me on an early proof-of-concept version of the BBE race simulator written in Kotlin when I was advisor on his Masters-thesis project (see \cite{stuart_2019}), and to James Hawkins, James Keen, and Roberto Lau-Soto for their comments on earlier versions of this paper and for conversations about their implementations of the BBE model described here.

\bibliographystyle{apalike}

\bibliography{../dc_bibliography}

\end{document}